# Brain Tumor Classification Using Deep Learning Technique - A Comparison between Cropped, Uncropped, and Segmented Lesion Images with Different Sizes

**Ali Mohammad Alqudah[1,*], Hiam Alquraan[1], Isam Abu Qasmieh[1], Amin Alqudah[2], Wafaa Al-Sharu[3]**

[1]Department of Biomedical Systems and Informatics Engineering, Yarmouk University, Irbid, Jordan
[2]Department of Computer Engineering, Yarmouk University, Irbid, Jordan
[3] Department of Electrical Engineering, Hashemite University, Zarqa, Jordan
*Corresponding Author; Email address: ali_qudah@hotmail.com; ORCID: https://orcid.org/0000-0002-5417-0043

## ABSTRACT

Deep Learning is the newest and the current trend of the machine learning field that paid a lot of the researchers' attention in the recent few years. As a proven powerful machine learning tool, deep learning was widely used in several applications for solving various complex problems that require extremely high accuracy and sensitivity, particularly in the medical field. In general, the brain tumor is one of the most common and aggressive malignant tumor diseases which is leading to a very short expected life if it is diagnosed at a higher grade. Based on that, brain tumor grading is a very critical step after detecting the tumor in order to achieve an effective treating plan. In this paper, we used Convolutional Neural Network (CNN) which is one of the most widely used deep learning architectures for classifying a dataset of 3064 T1 weighted contrast-enhanced brain MR images for grading (classifying) the brain tumors into three classes (Glioma, Meningioma, and Pituitary Tumor). The proposed CNN classifier is a powerful tool and its overall performance with an accuracy of 98.93% and sensitivity of 98.18% for the cropped lesions, while the results for the uncropped lesions are 99% accuracy and 98.52% sensitivity and the results for segmented lesion images are 97.62% for accuracy and 97.40% sensitivity.

**Key words:** Deep Learning, Brain Tumor, Brain Lesions, Cropped lesion, Uncropped lesion, segmented lesion.

## 1. INTRODUCTION

According to the World Health Organization (WHO) definition of a brain tumor, which has been reclassified in 2016, a brain tumor is a type of tumors that affect the central nervous system. In general, brain tumors are defined as a group of brain cells that grow in an abnormal way. Such type of tumors makes the brain tissue exposed to a size declination which leads to mass damage to the neural network of the brain which consequently disrupts the work of the brain [1, 2, 3, 4]. There are two types of brain tumors (like any type of cancer), namely cancerous or non-cancerous (benign and malignant tumors) and the types of brain tumors that generally occur based on the affected area are meningioma, glioma, and pituitary. Each type of these tumors has a certain level of malignancy. Glioma is a type of brain tumor that grows on the area of the glia tissues and spinal cord, meningioma is a type of tumor that grows on the area of the membrane (the area that protects the brain and spinal cord), while the pituitary tumor grows on the pituitary gland area [3, 5, 6].
Typically, the initial evaluation of brain tumors by oncologists usually performed using medical imaging techniques such as Magnetic Resonance Imaging (MRI) and Computed Tomography (CT) scans. These two modalities are widely used to produce highly detailed images of the brain structure and any changes can be noticed. However, if the physician suspects a brain tumor and they require more information about its type, a surgical biopsy from the suspected tissue (tumor) is necessary for a detailed diagnosis by the specialist. These different technologies in brain tissue imaging have increased over recent years for image contrast and resolution enhancement which allows the radiologist to identify even small lesions and therefore achieving higher diagnosis accuracy[7, 8,9, 10].
Combining (fusing) the images acquired from different imaging modalities and benefiting the recent advances of engineering technologies that enhance the accuracy of brain tumor detection in the field of artificial intelligence (AI) for computer vision applications, where AI can be integrated with these imaging modalities to build a computer-aided diagnosis (CAD) systems. Such systems can help physicians to increase the accuracy of cancer early detection. Recently, many artificial intelligence techniques such as an artificial neural network (ANN), support vector machine (SVM), and convolutional neural network (CNN) have been applied to classify and recognize brain tumors [11-15].

CNN represents the most recent advancement and the state of the art in the machine learning field, which is employed in the field of diseases diagnosis based on medical images, particularly the CT and MRI images. CNN has been recently widely used in the medical imaging classification and grading since it does not require preprocessing or features extraction before the training process [16, 17]. In general, CNN's are designed to minimize or canceling in sometimes the data pre-processing steps and usually are used to deal with raw images. CNN is consisting of many stacked layers in the following order: the input layer, convolution layer, RELU layer fully connected layer, classification layer, and output layer. The process of CNN is mainly consisting of and relies on two processes; the convolution, which is performed using trainable filters with the pre-determined specification that are tuned during the training phase, and the downsampling [18, 19].

In general, the machine learning applications for brain tumor classification can be divided into two groups: classifying brain MRI images to two main categories normal and abnormal and grading the abnormal brain MRI images into several types of brain cancer. The CNN has been used for detection and grading of brain tumors and it pays researchers' attention as a strong tool in diseases detection and classification field which in turn will increase the accuracy of the detected brain tumor grading and help physicians in determining a perfect treatment plan and therefore increasing the healing percentage [20 - 23].

This research was conducted to grade the brain tumor using two scenarios, cropped tumor lesions, and uncropped brain images. The model images are T1-Weighted contrast-enhanced brain tumor MRI images. These images are fed into a novel CNN architecture and trained to calculate the weight of networks. The results show that both cropped and uncropped have high accuracy, high sensitivity, and high specificity. The contributions of this research work can be summarized in the following points: A new CNN architecture has been used rather than using transfer learning techniques and pretrained CNN like Densenet201. Evaluating the model performance of the two images sets, cropped, uncropped, and segmented lesion images. The rest of the paper is organized as follows: In Section 2, we introduce a literature review of the current methods that are addressing the problem of brain tumor grading. Next, the material and methods used in this study are described in detail in section 3. Experimental results are presented in Section 4. Sections 5 and 6 provide the results discussion and finally the proposed techniques conclusions.

## 2. LITERATURE REVIEW

Recently, Machine learning (ML) and Deep Learning (DL) methods are widely been used for detection and grading brain tumors using different imaging modalities, especially those acquired using MRI. In this section, the most recent and related research works on the paper topic are presented. Mohsen, Heba, et al. [8] propose a system that combines discrete wavelet transform (DWT) features and deep learning (DL) techniques. They have used fuzzy c-mean method for segmenting the brain tumor, and for each detected lesion the DWT was applied to extract the features, where these features are fed into the principal component analysis (PCA) for feature dimension reduction and finally the selected features are then fed to deep neural networks (DNN). The results show that they achieve an accuracy rate of 96.97% and a sensitivity of 97.0 %. Widhiarso, Wijang, Yohannes Yohannes, and Cendy Prakarsah [10] presented a brain tumor classification system using a convolutional neural network (CNN) and Gray Level Co-occurrence Matrix (GLCM) based features. They extracted four features (Energy, Correlation, Contrast, and Homogeneity) from four angles (0°, 45°, 90°, and 135°) for each image and then these features are fed into CNN, they tested their methodology on four different datasets (Mg-Gl, Mg-Pt, Gl-Pt, and Mg-Gl-Pt) and the best accuracy achieved was82.27% for Gl-Pt dataset using two sets of features; contrast with homogeneity and contrast with correlation.

Seetha, J., and S. S. Raja [12] proposed a deep CNN based system for automated brain tumor detection and grading. The system is based on Fuzzy C-Means (FCM) for brain segmentation and based on these segmented regions a texture and shape features were extracted then these features were fed into SVM and DNN classifiers. The results showed that the system achieved a rate of 97.5% accuracy. On the other hand, Cheng, Jun, et al. [13] enhanced the performance of the brain tumor classification process using region of interest (ROI) augmentation and fine ring-form partition. They applied these enhancements to different feature extractions methods which are intensity histogram, GLCM, and the bag-of-words (BoW) where these features vectors are fed into a classifier. The experimental results showed that the accuracy enhanced from 71.39% to 78.18%, and 83.54% to 87.54%, and 89.72% to 91.28% for intensity histogram, GLCM, and BoW respectively. Sasikala, M., and N. Kumaravel. [17] proposed a genetic algorithm feature selection for feature dimension reduction of wavelet features set. The method is based on selecting optimal features vector that can be fed into the selected classifier such as an artificial neural network (ANN). The results show that the genetic algorithm selected only 4 of 29 features and achieved an accuracy of 98% using only the selected features. Khawaldeh, Saed, et al. [23] proposed a system for non-invasive grading of glioma brain tumors using a modified version of AlexNet CNN. The classification process was done using whole-brain MRI images and the labels of the images were at the image level, not the pixel level. The experimental results showed that the method achieved a reasonable performance with an accuracy of 91.16%. Sajjad, Muhammad et al. [24] proposed an extensive data augmentation method fused with CNN for brain tumor classification. The method used for multi-grade classification of brain tumors using segmented brain tumor MRI images. They used pretrained VGG-19 CNN architecture for classification using transferee learning and achieved an overall accuracy of values 87.38% and 90.67% for data before and after augmentation respectively. While Özyurt, Fatih et al. [25] combine the CNN with neutrosophic expert maximum fuzzy (NS-CNN) sure entropy for brain tumor classification. They used the neutrosophic set – expert maximum fuzzy-sure method for brain tumor segmentation then these images are fed to CNN to extract features and fed them to SVM classifiers to be classified as benign or malignant. They achieved an average success of 95.62%.

## 3. MATERIAL AND METHODS

The main aim and motivation behind this research paper are to provide a new CNN architecture for grading (classifying) brain tumors using T1-weighted contrast-enhanced brain MR images. Figure 1 Shows the Block diagram of the proposed methodology. In this section, the following sub-sections are discussed in detail; the used dataset, and the proposed methodology.

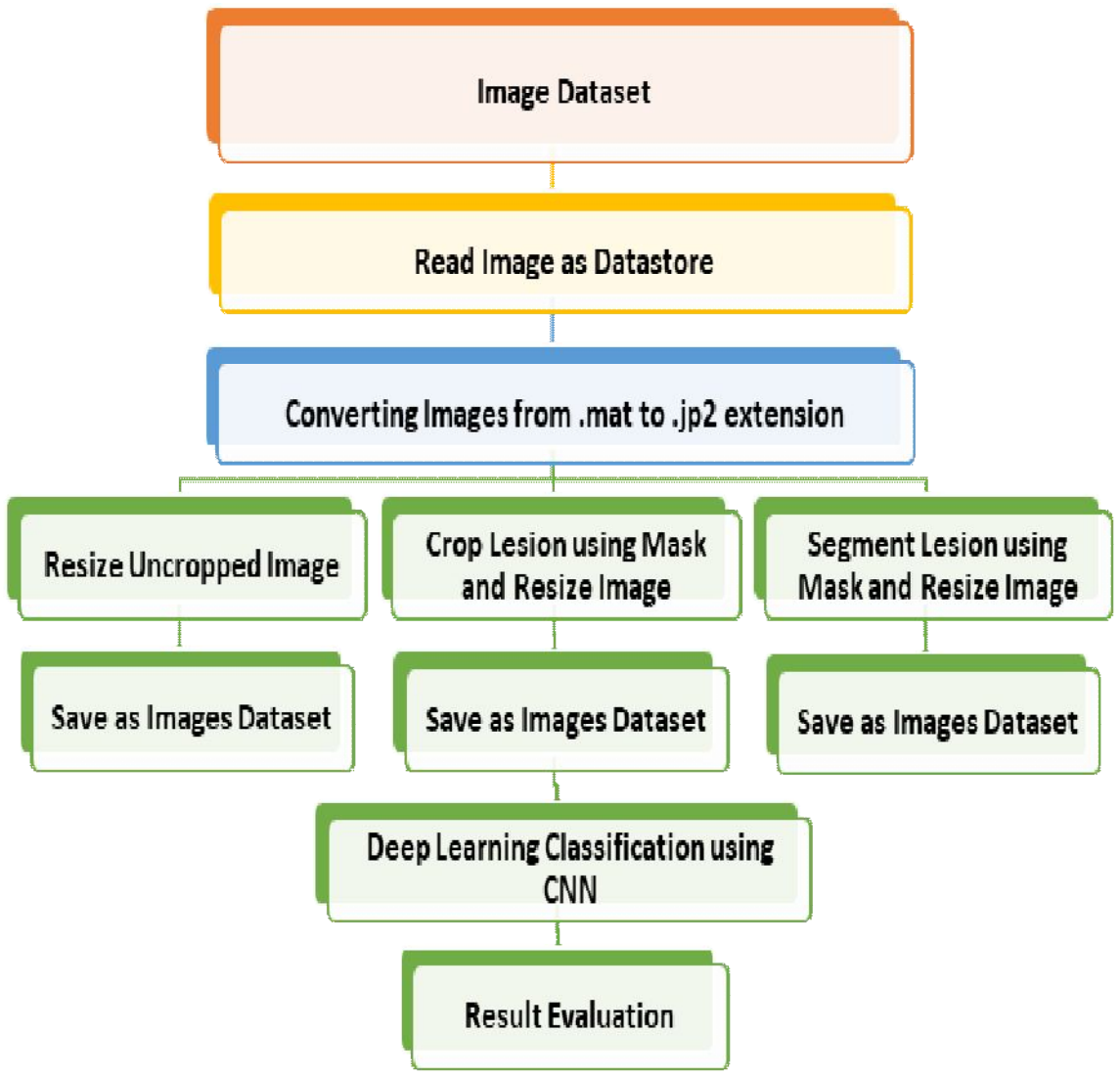

**Figure 1:** Block Diagram of the Proposed Methodology.

### 3.1 Dataset

In this paper, we used the brain tumor dataset proposed by Cheng, Jun, et al. [13] which is available online for free at https://figshare.com/articles/brain_tumor_dataset/1512427/5. The dataset contains a 3064 T1 weighted and contrast-enhanced brain MRI's images and includes three classes glioma, meningioma, and pituitary tumor. Table.1 lists the number of images of each class in the dataset. about each image in the dataset is fully described and a full information is provided; such as the patient id (PID), tumor mask, tumor border, and the class label, where the most important information after the class label is the lesion mask which is used to crop the tumor region of interest (ROI). Figure 2 shows a sample of cropped and uncropped lesions images from the used dataset.

**Table 1:** Summary of Used Image Dataset.

| Class | Number of Images |
|---|---|
| Glioma | 1426 |
| Meningioma | 708 |
| Pituitary Tumor | 930 |
| **Total** | **3064** |

### 3.2 Methodology

### 3.2.1 Convolutional Neural Network (CNN)

Convolutional neural networks (CNN) are currently the most widely used deep-feed forward neural networks that can treat different types of data inputs either 2D images or 1D signals. In general, CNN consists of many layers namely; input layer, convolution layer, RELU layer, fully connected layer, classification layer, and output layer [8, 26].

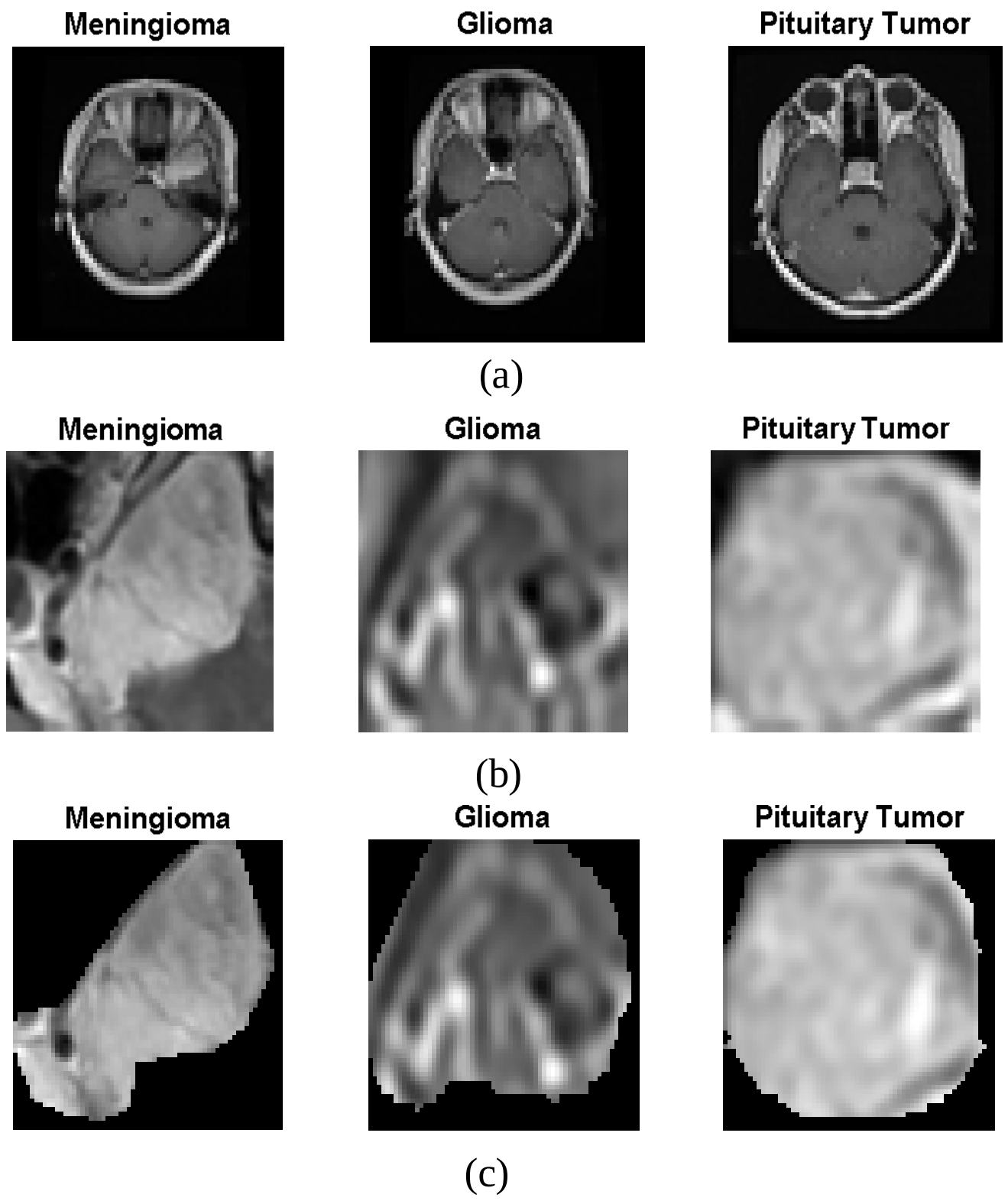

**Figure 2:** Sample from the used dataset; (a): Uncropped images, (b) Cropped images, and (c) Segmented images.

CNN is basically based on two processes; convolution using a trainable filter which has a pre-specified size, and weights that adjusted during the downsampling process in the training phase to achieve a high accuracy [10, 26]. In this research, the cropped and uncropped brain tumors images are stored as a database and three folders are created, each one consists of the images for specific class glioma, meningioma, and pituitary tumor. The database is partitioned into training and testing data, where 70% of the data is utilized in the training stage and the rest is used in the test stage. A new CNN architecture is employed in this paper. The next sections will explain the structure of the proposed CNN architecture.

### 3.2.2 Proposed CNN Architecture

In this paper, we have proposed and used a newly designed CNN architecture. The architecture consists of 18 layers to enable the classifier to grade the brain tumor effectively. This architecture was firstly provided by Alqudah [27] for OCT images classification. In this paper the architecture is modified and transferred to be applied on three different images dataset; cropped, uncropped, and segmented and the performance of the architecture was evaluated. Figure 3 illustrates the structure of the used CNN architecture.

### 3.2.3 Performance Evaluation

To evaluate the performance of the proposed CNN architecture in grading the brain tumor in both cases; cropped and uncropped image, the confusion matrix for all cases (cropped, uncropped, and segmented ) were generated, and a comparison between the CNN architecture outputs with its corresponding original image label was carried out based on these generated confusion matrices. In general, using these generated confusion matrices we can calculate the accuracy, sensitivity, precision, and specificity, to measure how precisely the brain tumor being graded [28].

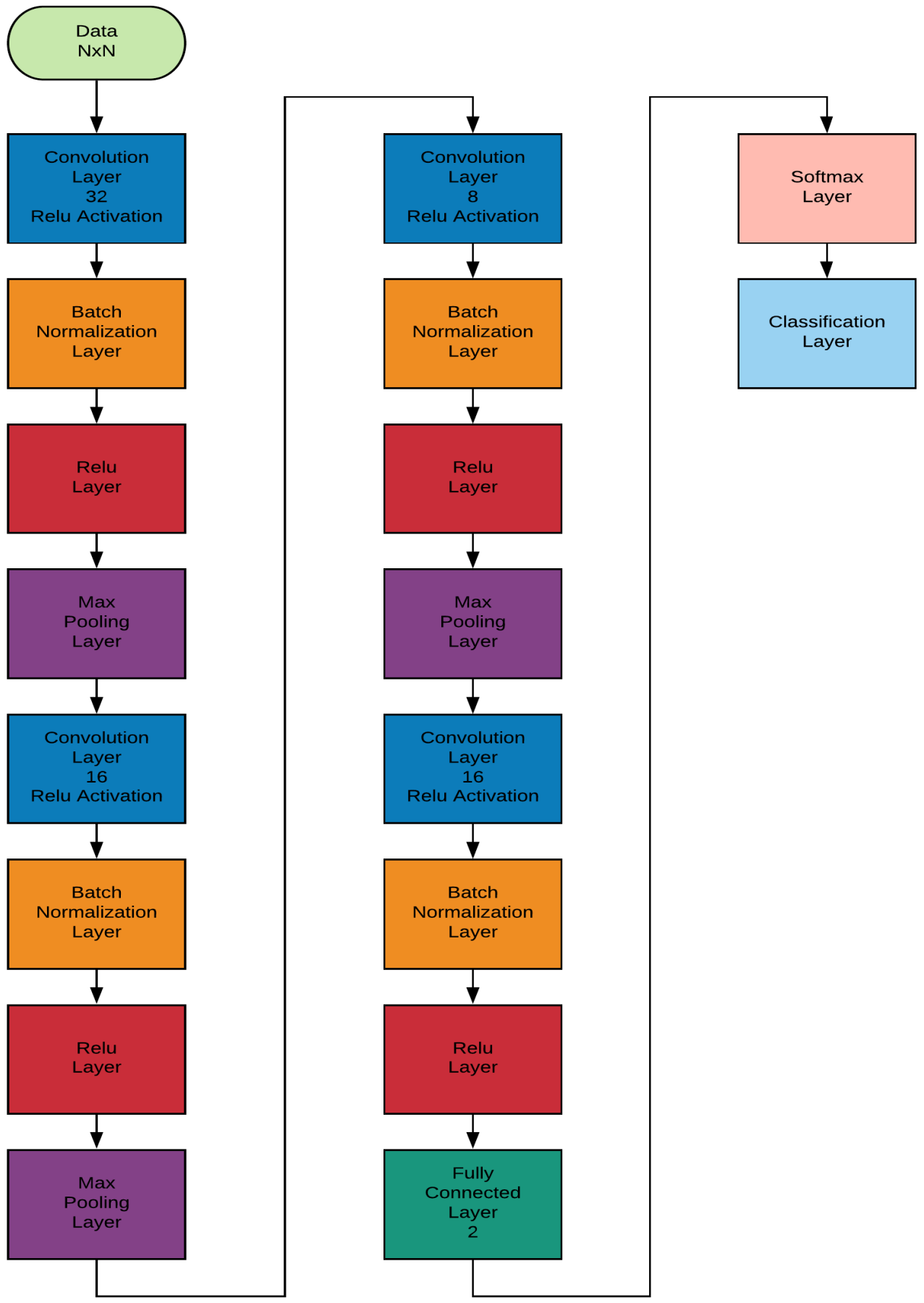

**Figure 3:** Proposed Convolutional Neural Network (CNN) Architecture.

From the generated confusion matrix, four statistical indices are calculated and used to evaluate the performance of the suggested classification system, namely true positive (TP), false positive (FP), false negative (FN), and true negative (TN) [28, 29] as follows:

$$Accuracy = \frac{TP+TN}{TP+FP+TN+FN} \quad (1)$$

$$Sensitivity = \frac{TP}{TP+FN} \quad (2)$$

$$Specificity = \frac{TN}{TN+FP} \quad (3)$$

$$Precision = \frac{TP}{TP+FP} \quad (4)$$

## 4. RESULTS

All experiments were executed using a desktop computer with Intel Core-I7 processor and 16 Gb RAM. Both image dataset cropped and uncropped were run with a minibatch size of 64, ADAM optimizer as optimizing method, and with learning initial rate of 10-3 which results in 1600 iterations. The dataset was divided into three subsets; training, validation, and testing with a percentage of 70%, 15%, and 15% respectively. The following sections report the results of the proposed two image datasets using the designed CNN architecture. Figure 4 shows the accuracy variation overtraining and validation process during the CNN training process.

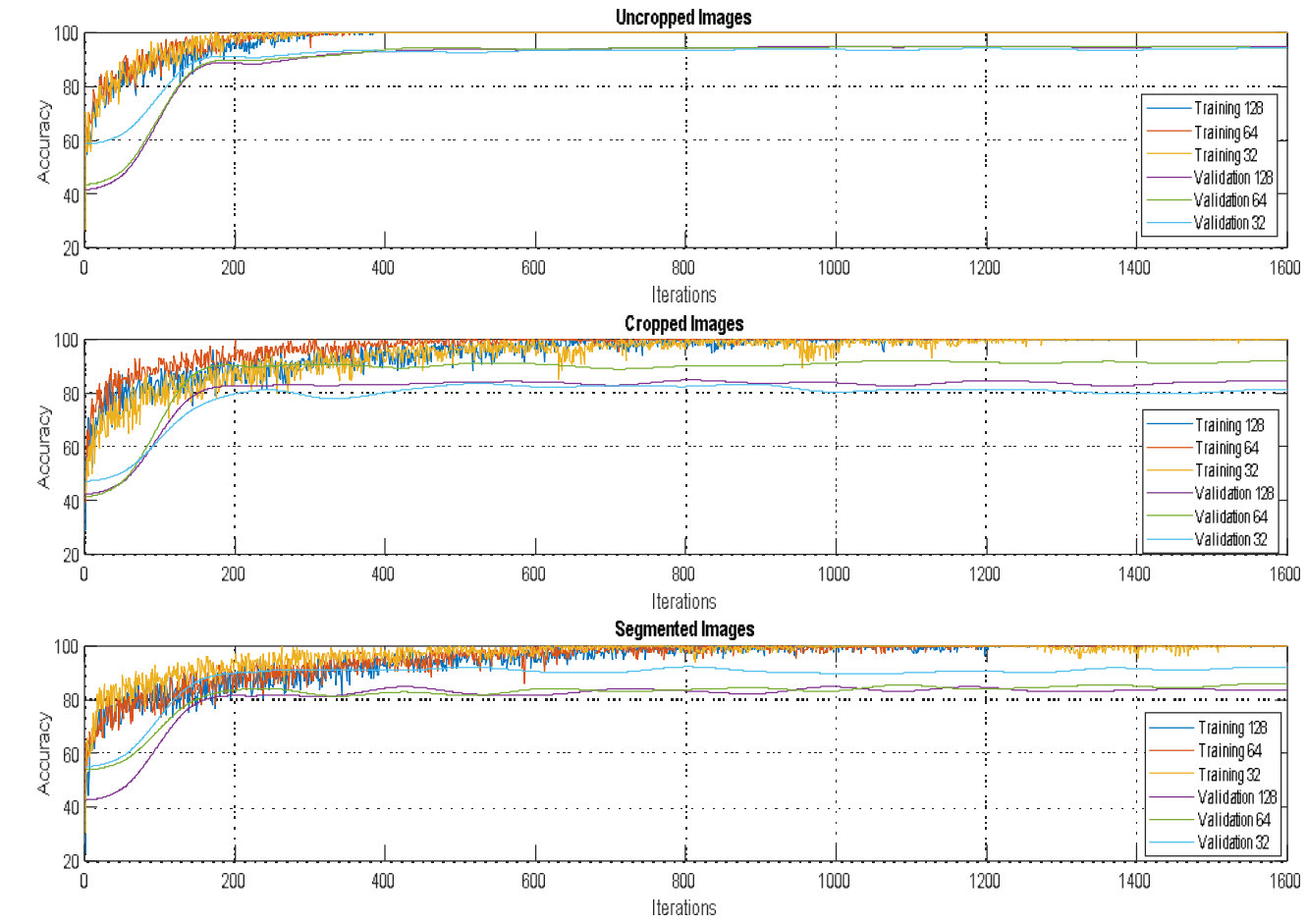

**Figure 4:** The Variation of Accuracy for Training and Validation of Three Cases over Training and Validation with Different Input Images Sizes During the CNN Training Process.

### 4.1 Results of Uncropped Lesion Images

The overall confusion matrices are shown in Figure 5. Based on these figures, we can note that the proposed system has efficiently graded the brain tumor with an accuracy rate values of 97.4%, 99.0%, and 99.2% for input images size of 128x128, 64x64, and 32x32 respectively.

### 4.2 Results of Cropped Lesion Images

The overall confusion matrices are shown in Figure 6. Based on these figures, we can note that the proposed system has successfully and efficiently had efficiently graded the brain tumor with accuracy rate values of 97.4%, 98.4%, and 96.9% for input images size of 128x128, 64x64, and 32x32 respectively.

### 4.3 Results of Segmented Lesion Images

The overall confusion matrices are shown in Figure 7. Based on these figures, we can note that the proposed system has successfully and has efficiently graded the brain tumor with an accuracy rate values of 97.5%, 97.6%, and 98.4% for input images size of 128x128, 64x64, and 32x32 respectively.

## 5. DISCUSSION

The dataset that has been used in this paper contains three types of brain tumors; Glioma, Meningioma and Pituitary tumors. In this work, an efficient automatic brain tumor classification is performed by using the proposed convolution neural network. Various manners have been applied to the dataset, such as segmented, cropped and uncropped tumors.

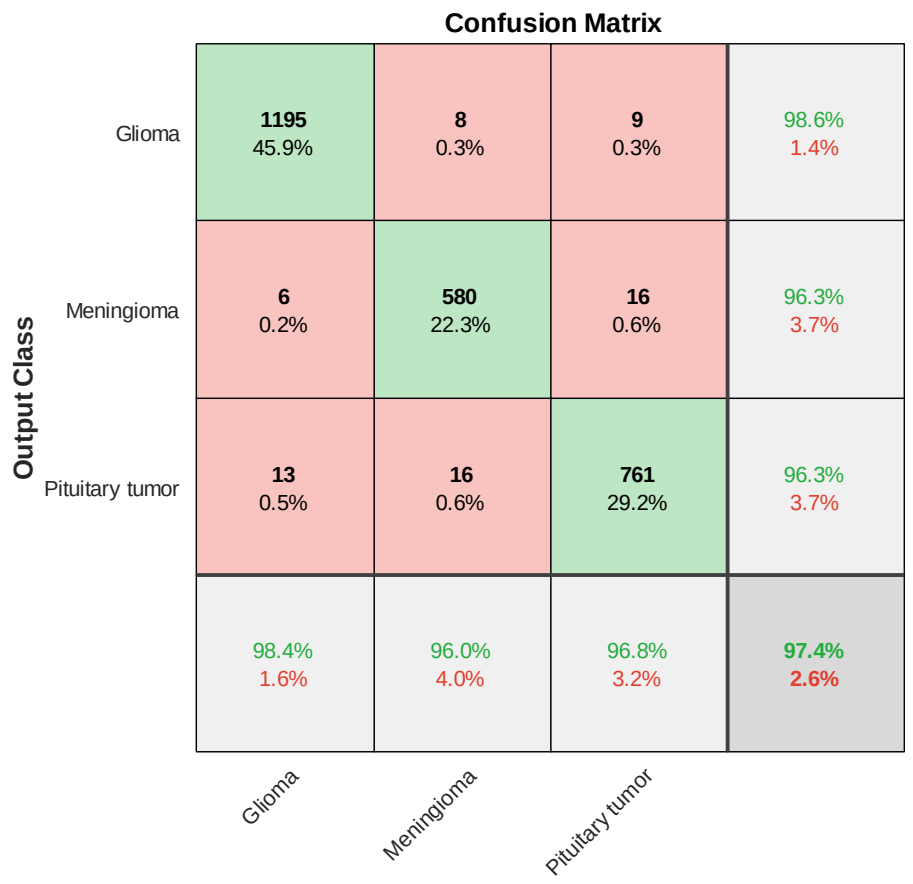

(a)

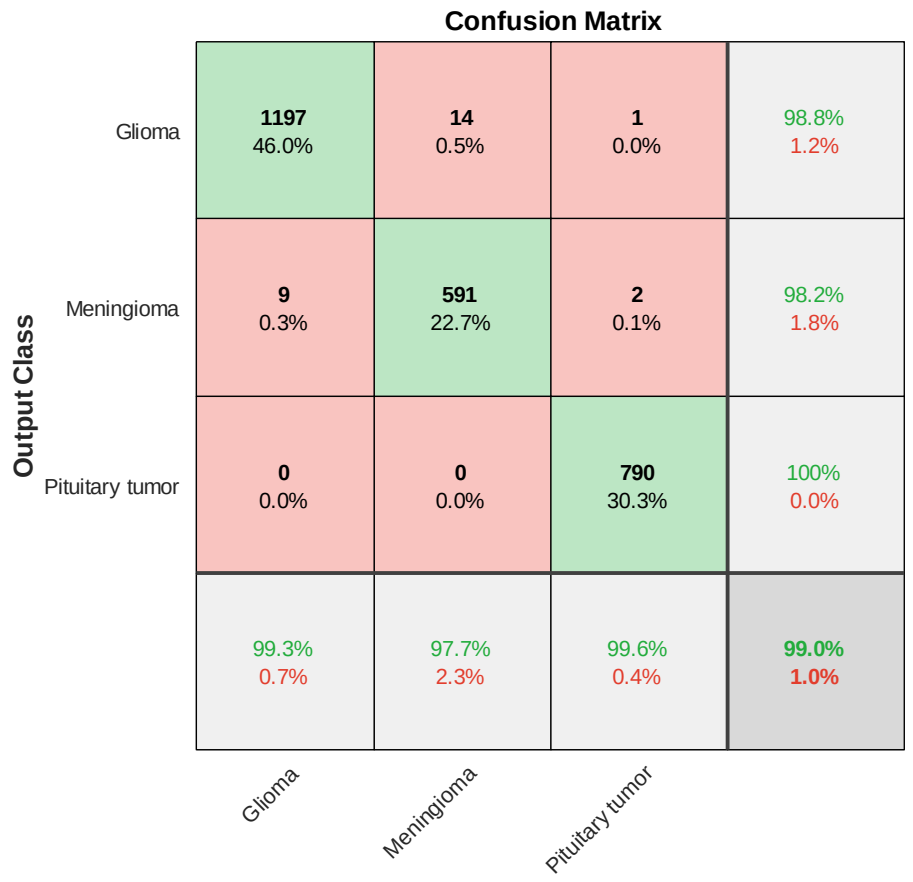

(b)

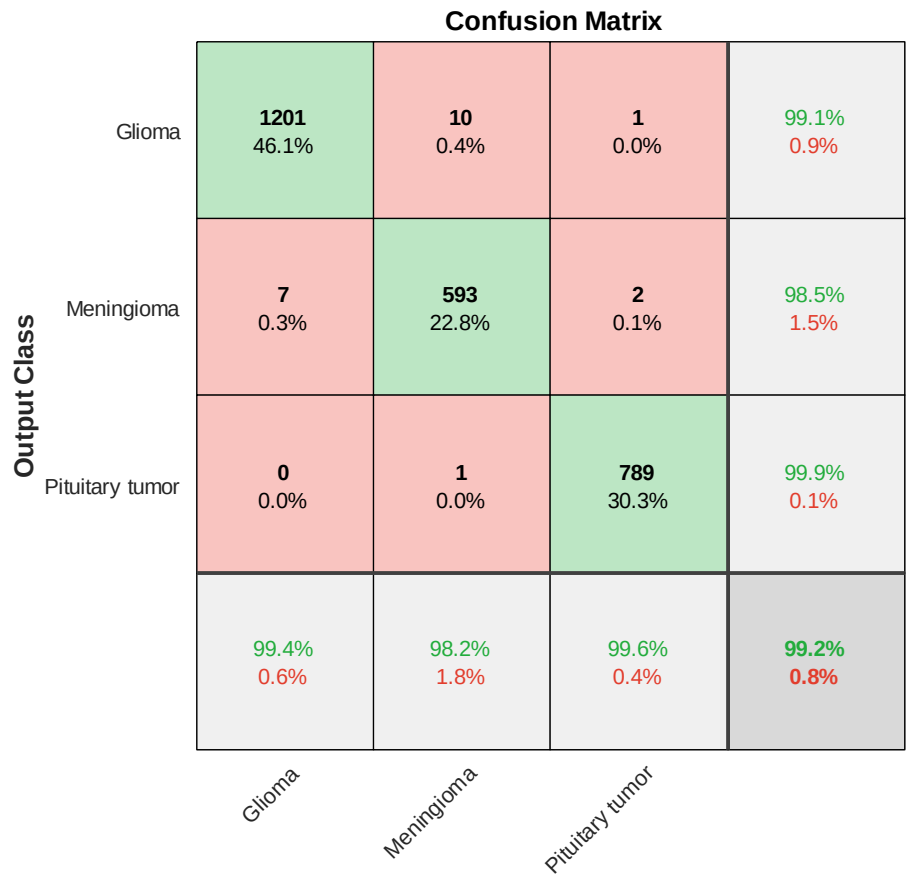

(c)

**Figure 5:** The Confusion Matrices for Uncropped Lesion Images; a: 128x128; b: 64x64; c:32x32.

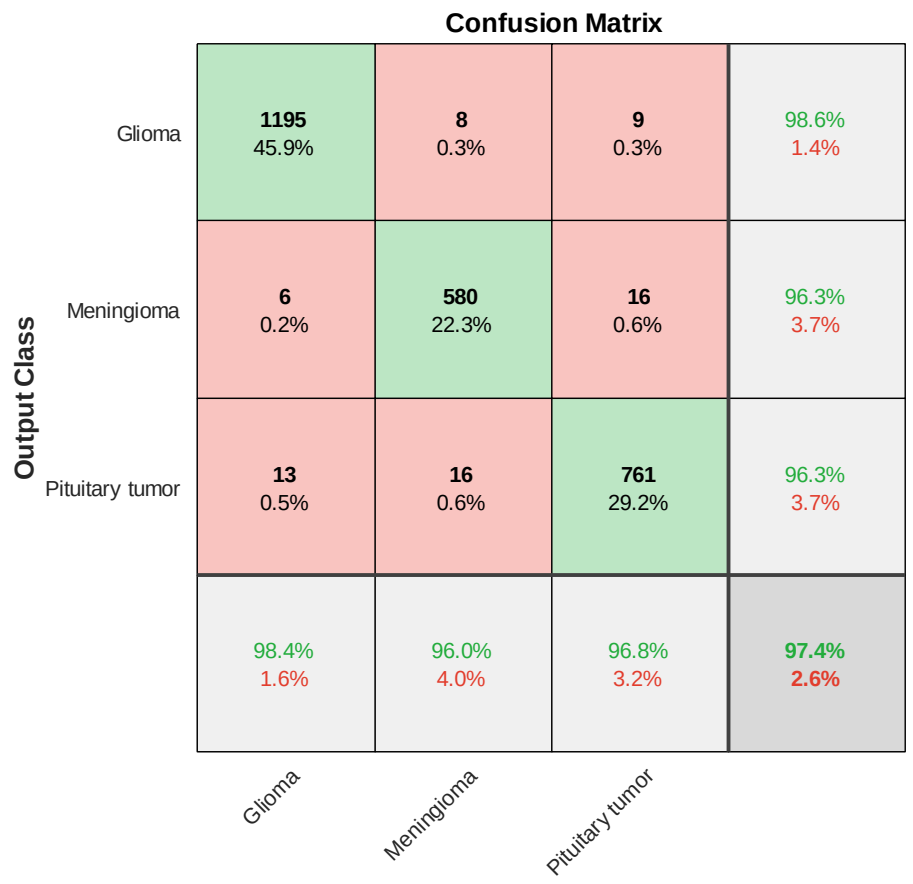

(a)

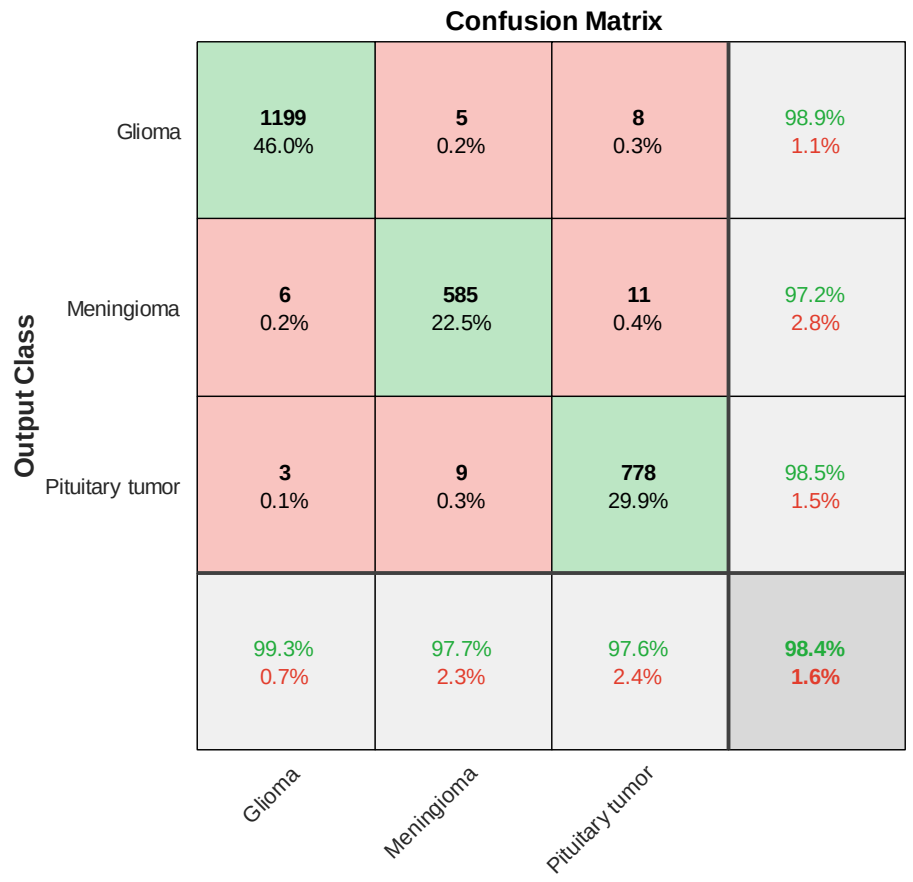

(b)

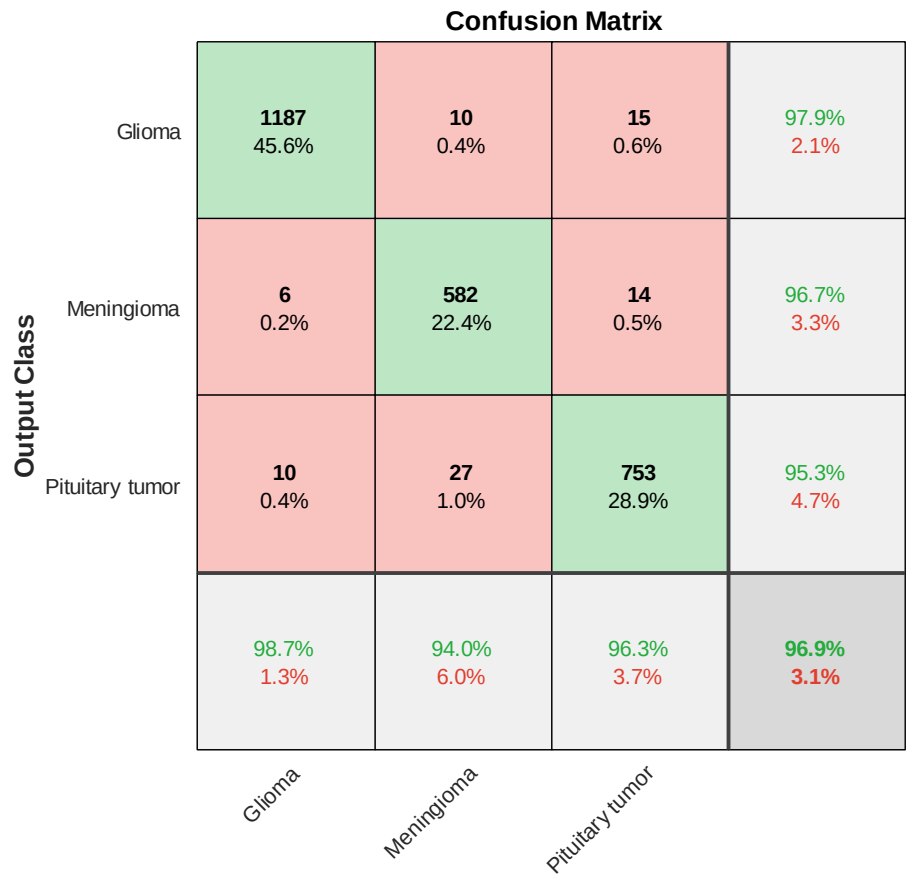

(c)

**Figure 6:** The Confusion Matrices for Cropped Lesion Images; a: 128x128; b: 64x64; c:32x32.

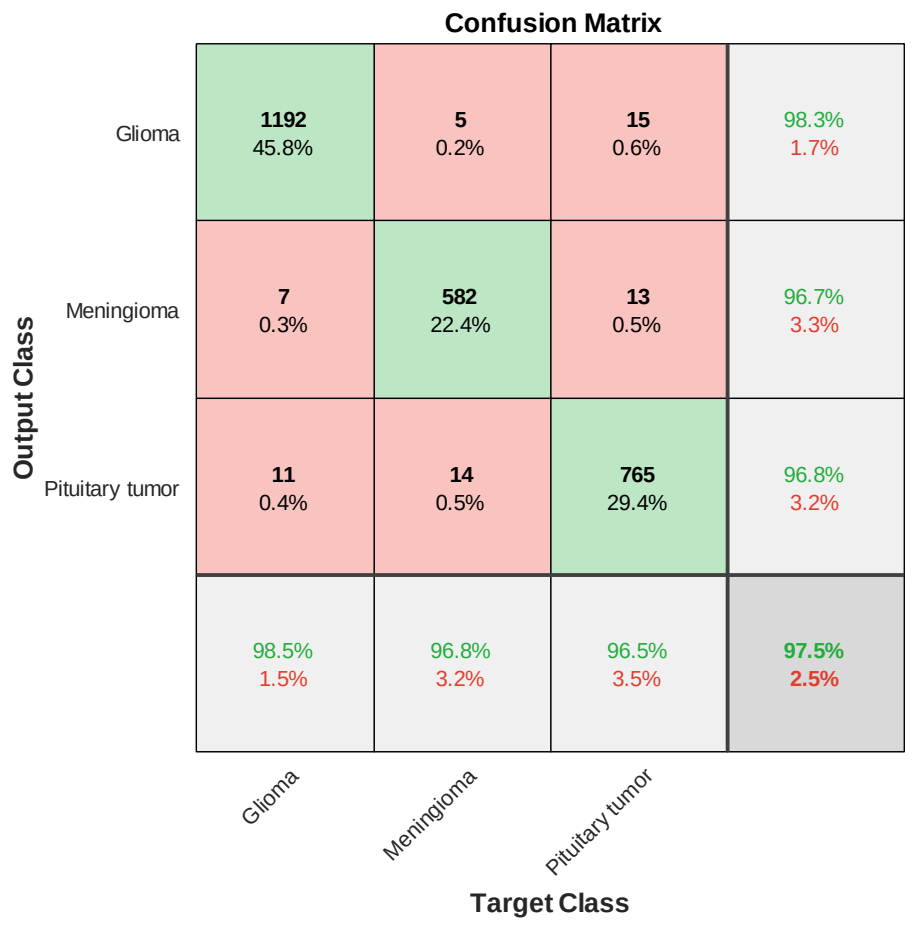

(a)

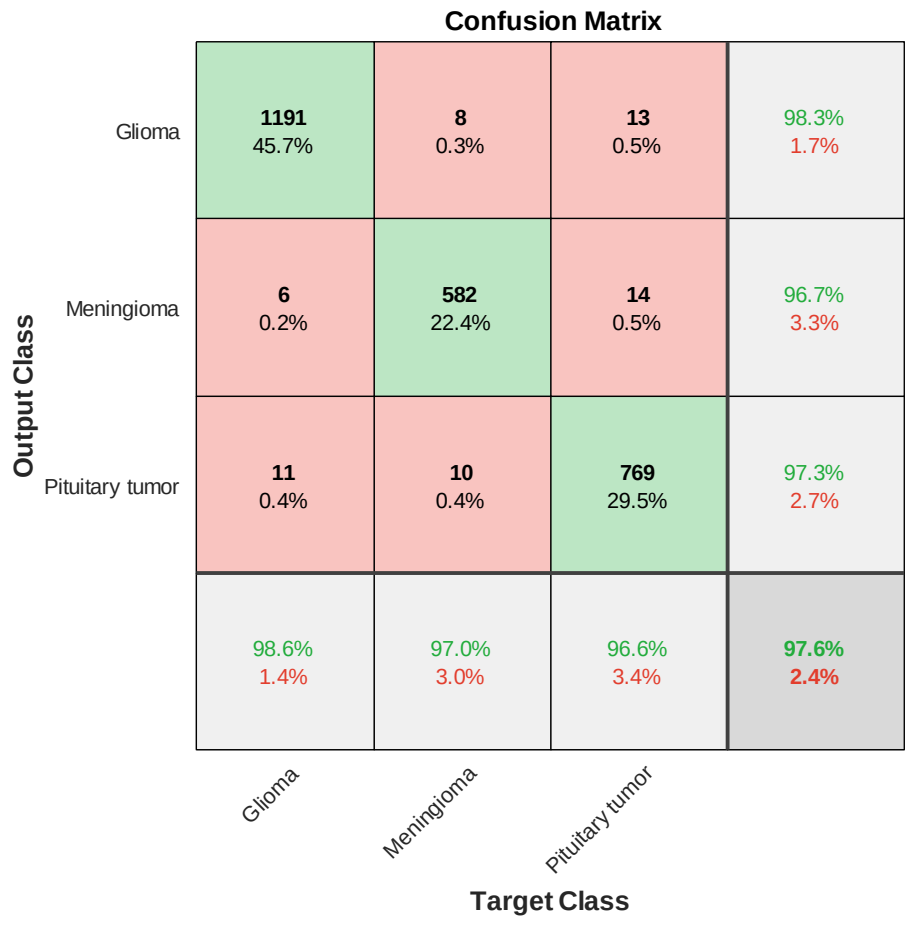

(b)

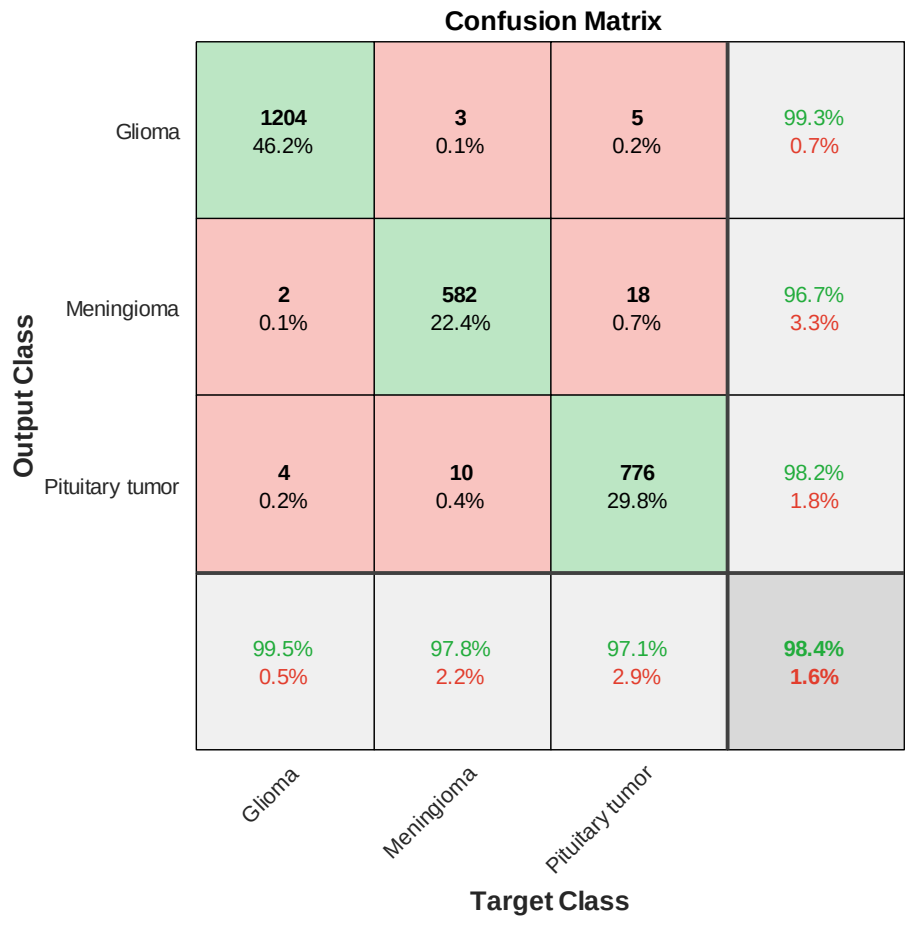

(c)

**Figure 7:** The Confusion Matrices for Cropped Lesion Images; a: 128x128; b: 64x64; c:32x32.

The statistical metrics (sensitivity, accuracy, specificity and precision) are calculated for each case. Figure 8, Figure 9, and Figure 10 respectively show the results for each size by utilizing the proposed CNN architecture, with the statistical performance of 99% for Accuracy, 99.64% for Specificity, 98.85% for Sensitivity and 98.98% for Precision in uncropped cases. On the other hand, using cropped images give a lower overall performance evaluation Accuracy as 98.40, Specificity as 99.19 %, Sensitivity as 98.18% and Precision as 98.19%. While segmented images give the lowest overall performance evaluation Accuracy as 97.62%, Specificity as 98.78 %, Sensitivity as 97.40% and Precision as 97.43%. As aforementioned, the uncropped images give the highest results when compared with cropped and segmented lesions. This comes from the cropping process which may discard some pixels around the lesion, which causes low discrimination to the type of the tumor when the results compared with the uncropped cases, which used all pixels and no discarding to anyone [30]. As shown in the presented results, the segmented cases give the lowest results, that is due to the texture color could be used for describing the lesion [31]. Typical color images composed of the three-color channels (RGB) red, green, and blue. In the segmented scheme we used the black color of the binary mask that surrounds the lesion, which is irrelevant to lesion color that is being classified, which leads to counting black in every image even it's not though presented in the lesion [32, 33]. For that reason, the efficiency of the proposed CNN in the segmented lesion is the lowest.

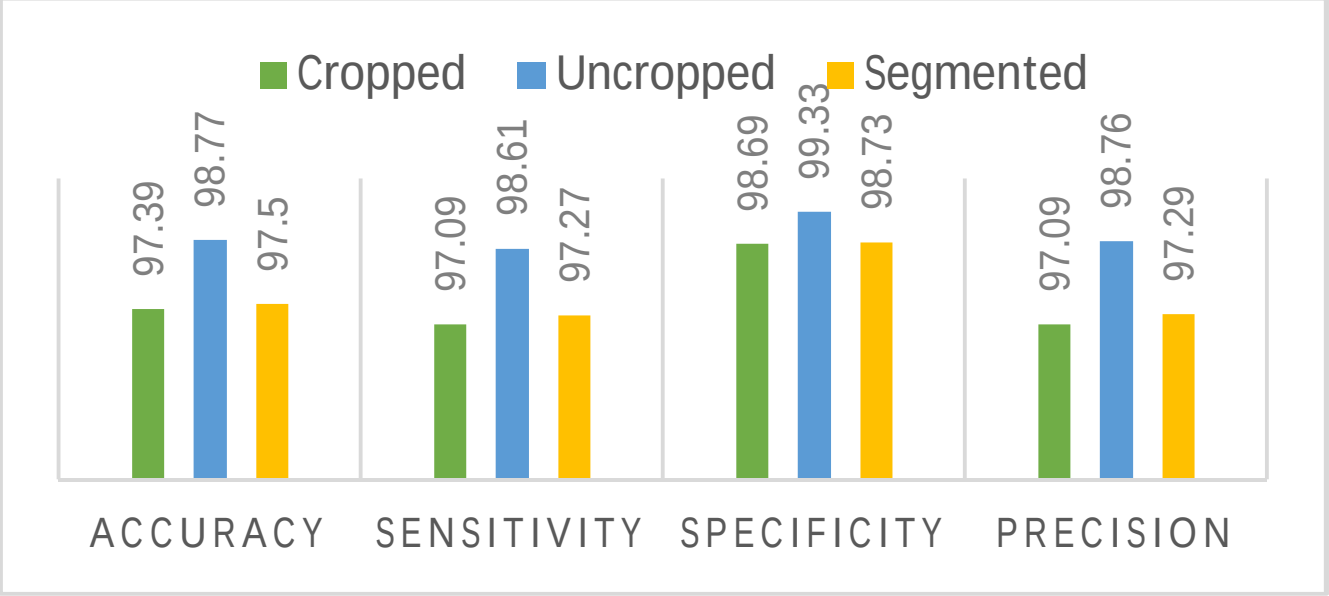

**Figure 8:** Performance Evaluation for Cropped, Uncropped, and Segmented Images Classification for Input Size of 128.

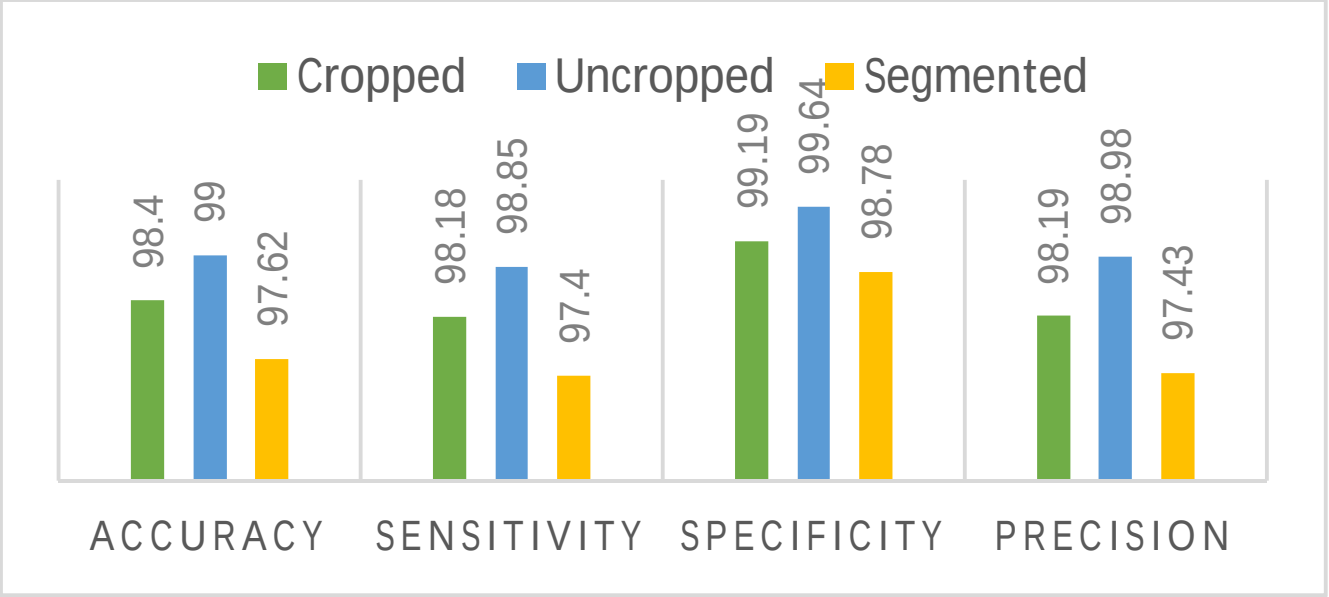

**Figure 9:** Performance Evaluation for Cropped, Uncropped, and Segmented Images Classification for Input Size of 64.

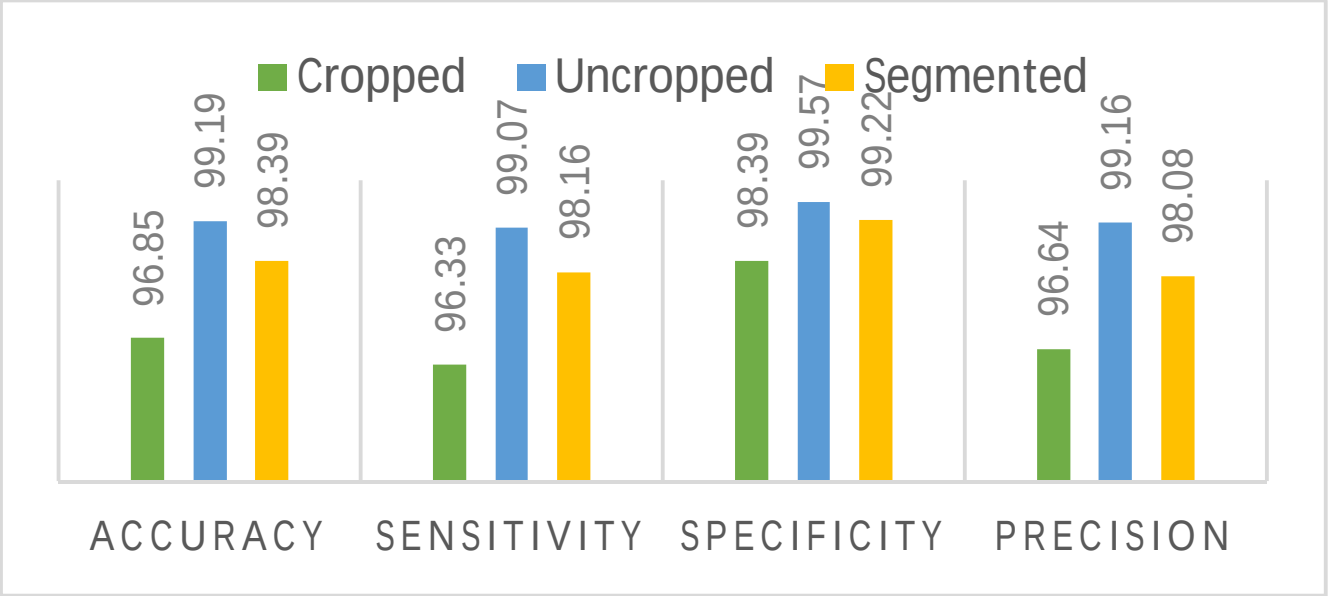

**Figure 10:** Performance Evaluation for Cropped, Uncropped, and Segmented Images Classification for Input Size of 32.

This paper focuses on multiclass (3 classes) classification problem to differentiate among glioma, meningioma and pituitary tumors, which form three prominent types of brain tumor. A comparison between the experimental results of the proposed system with the previous methods results in the literature is listed in Table 2. As shown in the table, most of the listed methods in the literature have achieved high recognition rates, greater than 90%, where the highest was 98%, and therefore, our method outperforms all the previous methods in the related studies with a good margin.

## 6. CONCLUSION

In this paper, we have presented a new convolutional neural network (CNN) architecture for automated grading (classification) of a brain tumor in three brain datasets; uncropped, cropped, and segmented region of interest (ROI). Our architecture succeeded in grading the brain tumor three classes with high performance in accuracy and sensitivity in all dataset cases; uncropped, cropped, and segmented. The system can significantly grade the tumor into three levels; meningioma, glioma, and pituitary tumor using T1 weight contrast-enhanced brain MR images. This architecture grading efficiency may even be further improved by including more brain MR images with different weights and with various contrast enhancement techniques to allow the architecture to be potentially more generalized and robust application for larger image databases.

**Table 2:** Comparison between the accuracy results in related work methods and the proposed method.

| Reference | Features Set | Classifier | Accuracy (%) | |
|---|---|---|---|---|
| [8] | DWT | DNN | 96.97 | |
| [10] | GLCM | CNN | 82.27 | |
| [12] | Texture and Shape | CNN | 97.50 | |
| [13] | Intensity Histogram GLCM and Bow | CNN | Intensity Histogram | 87.54 |
| | | | GLCM | 89.72 |
| | | | BoW | 91.28 |
| [17] | DWT | ANN | 98 | |
| [23] | CNN Based | CNN | 91.16 | |
| Proposed Cropped | CNN Based | CNN | 32x32 | 96.85 |
| | | | 64x64 | 98.93 |
| | | | 128x128 | 97.39 |
| Proposed Uncropped | CNN Based | CNN | 32x32 | 99.19 |
| | | | 64x64 | 99 |
| | | | 128x128 | 98.77 |
| Proposed Segmented | CNN Based | CNN | 32x32 | 98.39 |
| | | | 64x64 | 97.62 |
| | | | 128x128 | 97.50 |

## CONFLICT OF INTEREST

The authors confirm that there is no known conflict of interest associated with this publication.

## ACKNOWLEDGMENT

The authors would like to thank the owners of the Brain MRI images dataset to make the data available online. Also, the authors would like to thank the anonymous reviewers and the editor for their valuable comments.